\documentclass[reprint,superscriptaddress, amsmath,amssymb,aps,showkeys,showpacs,
twoside,nofootinbib ]{revtex4-2}
\usepackage[paperwidth=205mm,paperheight=290mm,top=17mm,bottom=25mm,
inner=17mm,outer=17mm,
twoside]{geometry}

\usepackage{cmap} 
\usepackage[T1,T2A]{fontenc}
\usepackage[utf8]{inputenc}
\usepackage[russian,english]{babel}
\usepackage{color}
\usepackage{graphicx}
\usepackage{dcolumn}
\usepackage{bm} 
\usepackage[unicode=true,colorlinks=true,linkcolor=magenta, urlcolor=blue, citecolor = blue,breaklinks]{hyperref}
\usepackage{multirow}
\DeclareGraphicsExtensions{.eps}

\newcount\issue
\newcount\Vol
\newcount\numb
\newcommand{\beq}{\begin{equation}}
\newcommand{\eeq}{\end{equation}}
\newcommand{\bea}{\begin{eqnarray}}
\newcommand{\eea}{\end{eqnarray}}

\newcommand{\bc}{\begin{center}}
\newcommand{\ec}{\end{center}}
\def\ds{\displaystyle}

\newcommand\blfootnote[1]{%
  \begingroup
  \renewcommand\thefootnote{}\footnote{#1}%
  \addtocounter{footnote}{-1}%
  \endgroup
}

\begin{document}

\title{Net-Baryon Number Probability Distribution \\
as an Indicator of Phase Transition} 

\def\addressa{NRC ”Kurchatov Institute” - IHEP, 142281 Protvino, Russia}
\def\addressb{Pacific Quantum Center, Far Eastern Federal University, 690922 Vladivostok, Russia}

\author{\firstname{R.N.}~\surname{Rogalyov}}
\affiliation{\addressa}
\author{\firstname{V.A.}~\surname{Goy}}
\affiliation{\addressb}


\begin{abstract}
The net-baryon number probability distribution 
and the related pressure and density are studied 
in the Cluster Expansion Model (CEM) and 
the obtained results are extrapolated to the domain of low temperatures. It is found that the rate of decrease
of the net-baryon number probability mass function 
diminishes with the temperature and, in 
the extensions of the CEM, falls to the level consistent with the phase transition at large baryon chemical potentials.
\end{abstract}

\par
\keywords{cluster expansion model, quantum chromodynamics, net-baryon number density, baryon chemical potential \\[5pt]}

\maketitle


\section{Introduction}\label{intro}

\blfootnote{{\it This is a preprint of the work published in "Moscow University Physics Bulletin", vol. 79} \copyright Allerton Press, Inc. 2024 (http://pleiades.online)}
In last decades, properties of 
strong-interacting matter at nonzero
temperatures and net-baryon number densities 
have received considerable study.
The QCD phase diagram was explored over a fairly large domain in the $\mu_B-T$ plane 
by carrying out experiments on collisions of heavy nuclei, as well as theoretically in lattice simulations of strong interactions or in QCD-inspired models.
The dependence of various thermodynamical 
quantities on the temperature $T$
and the baryon chemical potential $\mu_B$
obtained in these studies 
signals a rapid (though smooth) chiral 
crossover at $T=T_c=154(9)$~MeV and $\mu_B=0$,
the existence of the deconfinement phase
at $T>T_c$ and gives some evidence 
for the existence of the (currently, hypothetical) line of the 1st-order chiral phase transition at $\mu_B\sim 1$~GeV
and $T<T_c$. 

Unfortunately, straightforward
lattice-QCD simulations are impossible at 
$\mathrm{Re}\,\mu_B\neq 0$
due to the so called sign problem \cite{Gattringer_2016}.
For this reason, one employs some lattice 
results obtained at  $\mathrm{Re}\,\mu_B= 0$, 
$\mathrm{Im}\,\mu_B\neq 0$
and considers analytical continuation
in the complex plane of the parameter
$\ds \theta ={\mu_B\over T} = \theta_R+\imath \theta_I$. In this connection,
it should be remembered that the investigations
\cite{Roberge:1986mm} 
of analytical properties of the grand canonical
partition function $Z_{GC}(\theta)$ 
in the $\theta_I-T$ plane gave 
rise to the concept of the Roberge-Weiss 
temperature $T_{RW}$, which separates the domain 
$T> T_{RW}$ where 
the net-baryon number density $\rho(\theta)$
has discontinuities at $\ds \theta_I={2\pi n\over 3}$, $n\in \mathbb{Z}$ from the domain
$T< T_{RW}$ where 
$\rho(\theta)$ is an analytical function 
over some domain including the imaginary axis.
At all temperatures, $\rho(\theta)$ is a periodic function of $\theta_I$ with the period $\ds \theta_I={2\pi n\over 3}$.

Our attention in this work is concentrated on the 
net-baryon number density $\rho(\theta,T)$ and the additional pressure $p(\theta,T)$ caused by this density as well as their dependencies on $T$ and $\mu_B$. We try to stitch the patterns 
obtained from lattice-QCD data separately for  $T>T_{RW}$, $T_c<T<T_{RW}$, and $T\lesssim T_c$
in order to perform extrapolation to the domain
$T<T_c$, where a possibility of the
1st-order phase transition at large $\mu_B$ 
is under intensive current investigations.
Here we should draw attention to the 
generally accepted assumption that $\rho(\theta,T)$ 
has a discontinuity at the line of this transition
because it is not obvious that not only
the chiral condensate but also all other variables
are sensitive to this transition.
If the studies of $\rho(\theta,T)$ and 
related quantities do not provide evidence for this transition, its existence will not be ruled out:
such results can also be explained by
mutual independence of the physical 
mechanisms underlying the net-baryon number 
density and the chiral condensate.
In what follows we consider that the hypothetical chiral transition is related to a discontinuity in 
$\rho(\theta)$, that is, we search 
evidence for a phase transition of the Van der Waals type (VdW-type transition).

The main question discussed in this study
is as follows: do the results of naive 
extrapolation of the patterns of $\rho(\theta)$
behavior at $T\geq T_c$ to low temperatures
are consistent with the VdW-type transition?

An important quantity of phenomenological
significance associated with $\rho(\theta)$
is the net-baryon number probability distribution.
We study its behavior at $T>T_c$ on the basis 
of the models consistent with lattice data,
and argue that its slow decrease with the baryon number $n$
can be interpreted as a signal of the phase transition.

\section{Basic Definitions and formulas}\label{sec:quantities} 

The grand canonical partition function $Z_{GC}(\theta)$
and the canonical partition functions $Z_{C}(n)$
are related to each other by the formulas 
\bea
Z_{GC}(\theta) &=& \sum_{n=-\infty}^{\infty} Z_C(n) e^{n\theta}, \\ \nonumber 
Z_C(n) &=& {1\over 2\pi} \int_{-\pi}^\pi Z_{GC}(\theta)\Big|_{\theta_R=0} e^{\imath n \theta_I },
\eea 
$C$-parity conservation implies that 
$Z_C(n)=Z_C(-n)$.
It should be emphasized that 
the net-baryon number probability distribution 
at $\mu_B=0$, 
$\ds \bm{P}_n ={Z_C(n) \over Z_{GC}(0)} \;,$
contains all information on $\theta$-dependence:  
\beq 
{\cal P}_n(\theta)={Z_C(n)e^{n\theta} \over Z_{GC}(\theta)}
=\bm{P}_n e^{n\theta}\;{ Z_{GC}(0)\over Z_{GC}(\theta)}
\eeq 
where ${\cal P}_n(\theta)$ is the probability that the net-baryon number at given $T$ and  $\theta$ equals $n$.

We use the dimensionless pressure $\ds \hat p={p\over T^4}$
and net-baryon number density $\ds \hat \rho= {\partial \hat p\over \partial \theta} =\rho / T^3 = {n\over VT^3}$, where $n$ is the net-baryon number in the volume $V$ under consideration; we also use notation 
$\nu=VT^3$ for the parameter characterizing the number 
of occupied modes at temperature $T$ or the number 
of temperature-wave nodes in the volume $V$.

At $T>T_{RW}$ the imaginary part of the net-baryon number
density evaluated at imaginary chemical potential 
can be fitted by a low-degree polynomial,
\begin{equation}\label{eq:rho_vs_mu1}
-\imath\hat{\rho}_I = a_1\theta_I - a_3\theta_I^3 + a_5\theta_I^5 + ... 
\end{equation}

At $T<T_c$ $\hat \rho$ is best fitted by a trigonometric series
\beq\label{eq:fit_trig_series}
-\imath \hat\rho(\imath \theta_I) \simeq f_1 \sin \theta_I + f_2 \sin (2\theta_I) 
+ f_3 \sin (3\theta_I) +...  
\eeq
with a few (or even one) terms.

Numerical values of the fit parameters and more detailed description of the above basic formulas can be found in 
\cite{PhysRevD.95.094506,Rogalyov:2023pym}.

At $T_c<T<T_{RW}$ the best fit is provided by the 
Cluster Expansion Model (CEM) \cite{Vovchenko:2017xad}.

\section{ The Cluster Expansion Model as the base 
for the extrapolation} \label{sec:CEM}

For our purposes, the CEM,
can be defined  by fixing 
the coefficients $f^{CEM}_n$
in the Fourier expansion 
\beq\label{eq:Fourier_expansion_N}
 -\imath\,\hat \rho(\theta_I)\Big|_{\theta_R=0}=\sum_{n=1}^N f^{CEM}_n \sin\left({ n \theta_I}\right)  \;,
\eeq
of $\rho(\theta)$  
over the segment $\Big\{ \theta_R = 0; \ -\,\pi< 
\theta_I < \pi \Big\}$ 
as follows:
\beq\label{eq:aCEMexpr}
f_n^{CEM}\;=\;(-1)^{n+1}\;{b\, q^{n-1}\over n} 
\left[ 1+{3\over 4\pi^2 n^2}  \right] \; ,
\eeq
where the model parameters $b$ and $q$
depend on the temperature ($b>0$, $0\leq q\leq 1$).
Here we argue that the CEM, which agrees with 
lattice data well over the range $T_c<T<T_{RW}$ \cite{Vovchenko:2018zgt,Begun:2021nbf}, should not be used beyond this range.

At $q=1$ and $\ds b={16 N_F \over 81}$, 
$N_F$ is the number of light-quark flavors,
the CEM describes free-quark gas.

At $q=0$ and $\ds b={8\pi^2 \over 3+4\pi^2} 
\sqrt{\langle N_{B}\rangle \langle N_{\bar B}\rangle }$ the CEM describes the ideal 
Hadron Resonance Gas (HRG) 
provided that $\langle N_{B}\rangle $ and
$\langle N_{\bar B}\rangle$ are the average values 
of the baryons and anribaryons, respectively, in the 
ideal HRG model \cite{Braun-Munzinger:2011xux} (this being so, in formula (\ref{eq:aCEMexpr}) it is considered that $0^0=1$
so that the coefficient $f_1^{CEM}$ doesn't vanish).

In a simplified approach, when the 
free-quark gas model is assumed to be valid 
everywhere at $T>T_{RW}$ and the ideal HRG 
model --- at $T=T_c$, we 
suggest a simple dependence of $b$ and $q$
on the temperature:
\bea\label{eq:q_vs_T}
q &=& {T-T_c\over T_{RW}-T_c} \;, \\ 
b &=&  (b_2-b_1) {T-T_c\over T_{RW}-T_c} +b_1 \;,
\eea 
where $\ds b_1={8\pi^2 f_1\over 3+ 4\pi^2}$ and $\ds b_2={16 N_F\over 81}$.
Thus, $q$ runs through the entire range of 
its values allowed within the CEM (that is, from 1 to 0) 
as the temperature decreases from $T_{RW}$ to $T_c$. From a naive point of view, further decrease of the temperature suggests to consider
an extension of the CEM to the domain of negative values of $q$, namely, to $q\in [-1,0]$ (in the domain $|q|>1$ the series (\ref{eq:Fourier_expansion_N}) diverges.

Before considering such extension, we
study the behavior of $\hat \rho$,  $\hat p$, and $\bm{P}_n$ at $T_c<T<T_{RW}$ in more detail.

First we sum the series 
(\ref{eq:Fourier_expansion_N}),
perform analytical continuation 
from the imaginary axis to the  
complex $\theta$ plane, and
use the formula $\ds {\partial \hat p\over \partial \theta} = \hat \rho(\theta)$ in order
to obtain the expressions (see also \cite{Vovchenko:2019hbc,Begun:2021nbf})
\begin{widetext}
\beq\label{eq:rho_Li}
\hat \rho_{CEM}(\theta) = {b\over 2q }\left\{ \ln {1+q\exp(\theta ) \over 1+q\exp(\,-\;\theta )}
  + {3 \over 4\pi^2 }
\left[\mathrm{Li}_3\Big(-q e^{-\theta }\Big) - \mathrm{Li}_3\Big(-q e^{\theta }\Big) \right] 
  \right\} \ ,
\eeq

\beq\label{eq:p_Li}
\hat p_{CEM}(\theta) = - 
{b\over 2q}
\left\{ \mathrm{Li}_2
\Big(-qe^\theta \Big) 
+ \mathrm{Li}_2\Big(-qe^{-\theta}\Big) - 2 \mathrm{Li}_2(-q)  + {3\over 4\pi^2 }\left[\mathrm{Li}_4\Big(-q e^{\theta }\Big) + \mathrm{Li}_4\Big(-q e^{-\theta }\Big) - 2 \mathrm{Li}_4(-q) \right] \right\} \ ,
\eeq  

\end{widetext}
where the constant of integration is chosen so that
$\hat p(\theta=0)=0$.
We see that both $\hat \rho(\theta)$ 
and $\hat p(\theta)$ are analytical on  
the entire real axis, branch cuts lie 
on the line $\theta_I=\imath\pi$ \cite{PhysRevC.100.065202}. This gives evidence for the 
absence of the phase transitions at $T_c<T<T_{RW}$.

\begin{figure}[!htb]
\begin{center}
\includegraphics[width=1.\linewidth]{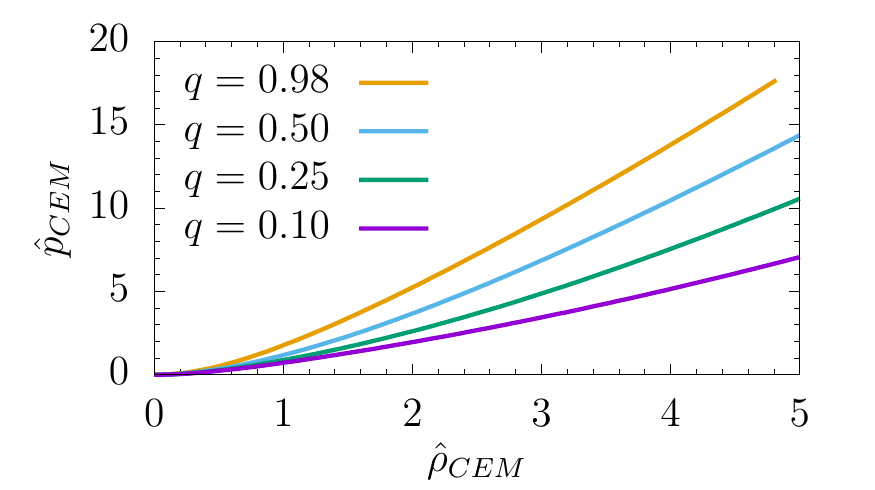}
\end{center}
\vskip -4mm 
\caption{The CEM isotherms obtained by formulas~(\ref{eq:rho_Li},\ref{eq:p_Li})}
\label{fig:Pressure}
\end{figure}

\noindent The evolution of the CEM isotherms in the $\hat \rho$--$\hat p$ plane with the decrease of the temperature is shown in Fig.\ref{fig:Pressure} ($q=0.98$ corresponds to $T_{RW}$, $q=0.1$ to a temperature slightly higher than $T_c$). Here 
and below,  in all numerical estimates we use $N_F=2$.

\section{Asymptotic behavior at $\theta\to\infty$ and low-temperature domain}

At $\theta \gg -\;\ln\,q$ one obtains the following asymptotic formula for the net-baryon number density:
\bea
\hat \rho &\simeq & {9 b\over 16 q} \left\{ (\theta+\ln\,q) + {(\theta+\ln\,q)^3\over 9\pi^2} \right\} \\ 
&& + {3+4\pi^2\over 4\pi^2}\;{1-q^2 \over q} e^{-\theta} + \underline{O}\Big(e^{-2\theta}\Big) \;. \nonumber 
\eea   
\noindent The canonical partition function
$\ds Z_C(n)=e^{-F(\hat \rho)/T}$
can be evaluated by the following procedure: 
the free energy as the function of the baryon number 
density is determined by the Legendre transform
\beq
F/T=\nu\Big( \hat\rho \theta_s(\hat \rho) - \hat p(\theta_s(\hat \rho))\Big)\;,
\eeq 
where $\theta_s(\hat \rho)$ can be found from the equation
$\ds {\partial \hat p \over \partial \theta}\Big|_{\theta=\theta_s}  =\hat \rho  \;.$
Performing this procedure and taking
the relation $\ds \hat \rho = {n\over \nu}$ into account, 
we arrive at the asymptotic behavior of the 
canonical partition functions at $n\to \infty$ 
\bea\label{eq:Zn_CEMfree_asympt}
Z_C(n) &\simeq&  \exp\Bigg[ -\,{9\over 4} \sqrt[3]{{3n^4 \over \nu N_F} }  + n\,\ln q  \\ 
&&  + \; {\pi\over 2}\,\sqrt[3]{9\pi N_F \nu n^2 } 
 + \underline{O}(1) \Bigg] \;.  \nonumber
\eea 
%

The case $q=1$ corresponds to the free-quark
gas, when the coefficients in formula (\ref{eq:rho_vs_mu1})  
are as follows: $\ds a_1 = {N_F\over 9}$, 
$\ds a_3 = {N_F\over 81\pi^2}$ and $a_k=0$ at $k\geq 4$.
We see that the leading asymptotic term 
in the free energy equals to that 
for the free quark gas and the pattern of 
asymptotic behavior depends only weakly on $q$.
The asymptotic regime (\ref{eq:Zn_CEMfree_asympt}) emerges when $n\gg \nu $. 

At $q\ll 1$ and $\hat \rho \ll -\ln q$ 
formula (\ref{eq:rho_Li}) corresponds to only
the first term in the series (\ref{eq:Fourier_expansion_N}), it 
tends to the relation $\hat \rho = 2\sqrt{\langle N_{B}\rangle \langle N_{\bar B}\rangle } \sinh(\theta)$
associated with the ideal HRG model.

As the temperature decreases below $T_c$, 
the high-order terms
in (\ref{eq:fit_trig_series}) become small so that
it is difficult to reliably evaluate high-order coefficients
$f_n$ on a lattice. For this reason, one is tempted 
to consider a straightforward extrapolation provided 
by formula of the type (\ref{eq:q_vs_T}); that is, 
to consider the CEM with $q<0$. 

In this case, the analysis of analytical properties of the net-baryon number density (\ref{eq:rho_Li}) reveals
branch-cut singularities of $\hat\rho(\theta)$ 
on the real axis of the $\theta$ complex plane
and the branch cuts along the real axis.
Since the branch cuts associated with $\hat\rho(\theta)$
in the baryon fugacity plane $\xi=e^\theta$
provide the lines of accumulation of Yang-Lee zeros
\cite{Lee:1952ig} and this branch cut lies on the 
positive semi-axis in the $\xi$ plane, 
we conclude that $Z_{GC}(\theta)=0$ for some values of 
$\theta: \theta_R>0, \theta_I=0$, which is unphysical. 
This being so, some of the canonical partition functions
$Z_C(n)$ must be negative.

Therewith, in the model with attractive interaction
associated with the VdW phase transition 
at low temperatures \cite{Vovchenko:2019hbc}, 
where the coefficients $f_n\sim q^n/n^{3/2}$
are positive and 
were found to be of the same sign,
the analytical properties of $\hat \rho(\theta)$
are quite similar. However, we can hardly evaluate 
high-density limit of the free energy using the Legendre
transform as well as the asymptotic behavior of $Z_C(n)$ 
at $n\to \infty$ precisely because 
of these branch-cut singularities.

Possible solution of this paradox is as follows.
The problem of unambiguous determination 
of the canonical partition functions $Z_C(n)$
from the known function $\hat p(\theta)$ (or $\hat \rho(\theta)$) is tightly connected with the 
problem of moments in probability theory:
the coefficients of the Maclaurin series for $\hat p(\theta)$  represent the cumulants of the net-baryon number probability distribution $\bm{P}_n$. It is well known~\cite{akhiezer2020classical} that the probability density function on an infinite-range interval cannot
be unambiguously restored from known cumulants 
when it decreases sufficiently slow ( $\bm{P}_n
> Ae^{-cn}$ for all $A$ and $c$ at $n\to \infty$). 
For example, a decrease like $\bm{P}_n\simeq \exp(-\alpha\sqrt{n^2 +a^2})$ would result only in the pressure being defined only in a limited $\theta$ range, whereas 
a decrease like $\bm{P}_n\simeq \exp(-\alpha\sqrt[4]{n^2 +a^2})$ would result in an ambiguity of reconstruction
of $\bm{P}_n$ from the respective cumulants.

\begin{figure}[!htb]
\begin{center}
\includegraphics[width=0.95\linewidth,trim=15mm 90mm 15mm 80mm,clip]{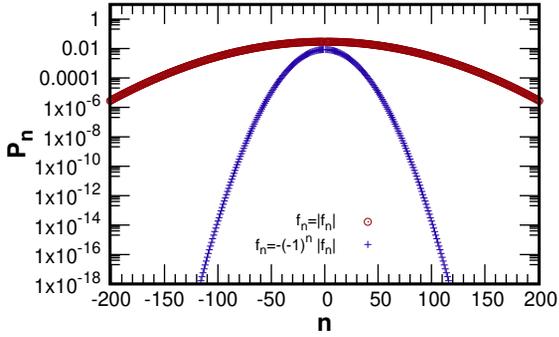}
\end{center}
\vskip -4mm 
\caption{The probabilities $\bm{P}_n$ {\it vs.} the net baryon number. Upper curve ($q=-0.4$) shows hypothetical behavior at $T<T_c$, lower curve ($q=0.4$)
--- at $T_c<T<T_{RW}$.}
\label{fig:EVMP}
\end{figure}

Fig.\ref{fig:EVMP} illustrates that positive coefficients 
$f_n$ lead to a significantly slower decreasing  of 
$\bm{P}_n$ than similar coefficients with alternating  
sign. The former characterizes the extension of the CEM to
negative values of $q$, the latter -- the CEM in the range of its validity $T_c<T<T_{RW}, 0<q<1$.

The above considerations allow us to formulate
a hypothesis that the temperature range associated with 
a phase transition at large $\mu_B$ is characterized 
by rather slowly (slower than $e^{-cn}$) decreasing 
net-baryon number probability distributions $\bm{P}_n$. 
This being so, the theory of the phase transitions 
is tightly related to the indeterminate problem of moments.


\section{CONCLUSIONS}

We have studied the dependence of the 
pressure and density of strongly interacting matter 
on the baryon chemical potential over the range
$T_c<T<T_{RW}$ in the framework of the CEM.
It has been shown that the asymptotic behavior
of the net-baryon number probability distribution 
$\bm{P}_n$ at $n\to \infty$ in the CEM agrees well
with that of the free-quark gas model, 
though it changes to the behavior $\bm{P}_n\sim 
\exp\big(-n\ln n\big)$ inherent in the ideal HRG model,
as the temperature drops to $T_c$.

The CEM results have been used for an 
extrapolation to low temperatures, where 
the extension of the CEM by considering the 
extended range of the parameter $q\in [0,1] \longrightarrow 
q\in [-1,1]$ has been studied, such extension is similar 
to the trivirial model \cite{Vovchenko:2019hbc} associated with the sought-for VdW-type phase transition at low $T<T_c$ and $\theta \gg 1$. In such models $\bm{P}_n$ shows
much slower decrease as compared with the CEM or the free-quark gas.

We have formulated the problem of negative probabilities $\bm{P}_n$ that arises at $q<0$ in the CEM or in the trivirial model when using the Lee-Yang approach.
The way to obviate this problem is to study ambiguities 
of reconstruction of $\bm{P}_n$ from the pressure
as well as the connections between the problem of moments 
in probability theory and the theory of phase transitions.

\section*{FUNDING}
This work is  supported by RSF (Project No. 23-12-00072, \href{https://rscf.ru/project/23-12-00072/}{https://rscf.ru/project/23-12-00072/}).

\section*{CONFLICT OF INTEREST}
The authors declare that they have no conflicts of interest.

\nocite{*}


\end{document}